\documentclass[a4papper]{jpconf}
\usepackage{graphicx}
\usepackage{enumerate}
\begin{document}
	\title{Measurement of the deuteron beam polarization at internal target at Nuclotron for DSS experiment}

	\author{Ya T Skhomenko$^{1,2}$, V P Ladygin$^1$, Yu~V~Gurchin$^{1}$, A~Yu~Isupov$^{1}$, M~Janek$^{3}$,  J-T~Karachuk$^{1,4}$,
A~N~Khrenov$^{1}$, P~K~Kurilkin$^{1}$, A~N~Livanov$^{1}$, S~M~Piyadin$^{1}$, S~G~Reznikov$^{1}$,  A~A~Terekhin$^{1}$, 
A~V~Tishevsky$^{1}$,  A~V~Averyanov$^{1}$,  A~S~Belov$^5$,  E~V~Chernykh$^{1}$,  D~Enache$^4$,
V~V~Fimushkin$^1$, D~O~Krivenkov$^{1}$}

\address{$^1$ Joint Institute for Nuclear Research, Dubna, Russia}
\address{$^2$ Belgorod State National Research University, Belgorod, Russia}
\address{$^3$ Physics Department, University of \v{Z}ilina, \v{Z}ilina, Slovakia}
\address{$^4$ National Institute for R\&D in Electrical Engineering ICPE-CA, Bukharest, Romania}
\address{$^5$ Institute of Nuclear Physics, Moscow, Russia}

	\ead{skhomenko@jinr.ru}

\begin{abstract}
The current deuteron beam polarimetry
at Nuclotron is provided by the Internal Target polarimeter based  
on the use of the asymmetry in $dp$- elastic scattering at large angles in the cms at 270 MeV.
The upgraded deuteron beam polarimeter has been used obtain the vector and tensor polarization during 2016/2017 runs
for the DSS experimental program. The polarimeter has been used also for tuning of the polarized ion source parameters 
for 6 different spin modes.
\end{abstract}

	\section{Introduction}
The study of the spin structure of two-nucleon  and three-nucleon short-range 
correlations via the measurements of the polarization observables in the deuteron induced 
reactions  is main goal of the DSS project at Nuclotron\cite{dss1,dss2,dss3}. 
The high precision  polarimetry of the deuteron and proton beams is important for these 
investigations. 

	The goal of the present article is to report  new results on the measurements of the vector and tensor beam polarizations
using upgraded polarimeter based on the asymmetry measurements in $dp$- elastic scattering at 270 MeV \cite{ITS_polarimeter} at the Internal Target Station (ITS)  at Nuclotron.   These measurements were performed during the DSS experiment on the study of the 
vector $A_y$, tensor $A_{yy}$ and $A_{xx}$ analyzing powers in $dp$- elastic scattering at large transverse momenta \cite{ladygin_dspin2017}.

\section{Deuteron polarimeter at ITS}
	 
Efficient polarimetry can be achieved even at relatively low beam intensity with using a thin solid target inside the inner ring of the accelerator. The luminosity can be increased significantly due to multiple beam passage through the interaction point and the use of a correctly configured internal target trajectory. Therefore, the internal beam polarimeter with a very thin target may have approximately the same efficiency as the extracted beams polarimeters.

The polarimeter based on the use of  $dp$- elastic scattering at large angles 
($\theta_{\rm cm}\ge 60^{\circ}$) at 270 MeV \cite{ITS_polarimeter}, where precise data on analyzing 
powers \cite{kimiko}-\cite{suda_nim} exist, has been developed at internal target 
station (ITS) at Nuclotron\cite{ITS}. The accuracy of the
determination of the deuteron beam polarization achieved with this method is 
better than 2\% because of the values of the analyzing powers were obtained for the polarized 
deuteron beam, which absolute polarization had been calibrated via the 
${\rm ^{12}C}(d,\alpha){\rm ^{10}B^*[2^+]}$ reaction\cite{suda_nim}.

Deuteron beam polarimeter  \cite{ITS_polarimeter} is placed in the Nuclotron ring. 
It consists of a spherical scattering chamber and system change targets that can be set six different targets.
A detector support
with 39 mounted plastic scintillation counters is placed downstream the ITS spherical chamber. Each
plastic scintillation counter was coupled to a photo-multiplier tube Hamamatsu H7416MOD. 
Eight proton detectors were installed for left , as well as for right
  and up, but due to space limitation -- only four for down.  The angular span
of one proton detector was 2$^\circ$ in the laboratory frame, which corresponds to
$\sim$4$^\circ$ in the cms.
Three deuteron detectors are placed at scattering
angles of deuterons coinciding kinematically with the protons. Only one deuteron
detector can cover the solid angle corresponding to four proton detectors placed down. 
In addition, one pair of detectors  is placed to register two protons from quasi-elastic $p-p$ scattering at $\theta_{pp}$=90$^\circ$ in the cms in the horizontal plane. 
The scattered deuterons and recoil protons at 270 MeV were detected in kinematic coincidence
over the cms angular range of 65--135$^\circ$ at eight different angles, defined by the positions of the proton detectors.


The VME (Versa Module Eurocard) based data acquisition system is used for the data taking from
scintillation detectors \cite{vme}.  The signals from the detectors  are fed in 16-channel TQDC-16 charge-time-digital 
converters via commutator bar.
TQDC-16 module allows to measure the amplitude and
time appearance of the signal simultaneously. 
The hardware of the DSS VME system consists of 4 TQDC-16 modules, trigger module TTCM and  VME controller.
There is a possibility to tune the first-level trigger using logic of trigger   and   TQDC-16  modules. 
This system has been significantly upgraded recently \cite{isupov_dspin2017}.   

Newly developed multichannel high-voltage power supply system (Wiener MPod) \cite{skhomenko} is used to provide the power 
for about 70 scintillation detectors  based on Hamamatsu photomultipliers.  
 
\section{Experiment at ITS} 

The internal target station (ITS) setup is well suited for study of the energy dependence of
polarization observables for the deuteron-proton elastic scattering and deuteron 
breakup reaction with the detection of two protons at large scattering angles.  
For these purposes the CH$_2$-target of 10 $\mu$m thick is used for the measurements.
The yield from  carbon content of the CH$_2$-target is estimated in separate measurements using several twisted 8$\mu$m
carbon wires. 
The measurements were performed using internal target station at Nuclotron \cite{ITS} with new   
control and  data  acquisition system \cite{ITS_DAQ}. 

New source of polarized ions (SPI) \cite{NewPIS} has been used to provide 
polarized deuteron beam. In the current experiment the spin modes with the maximal ideal values 
of ($P_z$,$P_{zz}$)= (0,0), (-1/3,-1) and (-1/3, +1) were used. 
The deuteron beam polarization has been measured at 270 MeV \cite{ITS_polarimeter}.

The DSS data taking was separated on 3 parts: November 2016, December 2016 and February 2017.
The deuteron beam polarization measurements were performed at 270 MeV before each energy studied in the range of 400-1800 MeV \cite{ladygin_dspin2017}. 
The beam polarization measurements were performed also in the end of each part of the experiment. 
The polarimeter \cite{ITS_polarimeter}  has been used to tune other spin modes of SPI.

The software for data analysis was developed using the ROOT package in C++.
The $dp$- elastic scattering  events at 270 MeV were selected using correlation of the energy losses 
and time-of-flight difference  for deuteron and proton detectors.  
The measurements were performed using CH$_2$ target only. 
The carbon contamination was measured to be less than 0.5\%. 

The precise data on the deuteron analyzing  powers at 270 MeV \cite{kimiko}-\cite{suda_nim} were used 
to get the polarization values at several angles \cite{ITS_polarimeter}. 
Assuming that the $Y$-axis is a symmetry axis ($\beta$=90$^{\circ}$, $\varphi$ =0$^{\circ}$) one can calculate $P_z$ and $P_{zz}$
using the normalized $dp$-elastic scattering events and analyzing powers known \cite{ITS_polarimeter}.
The values of the beam polarization for different spin  have been obtained  as weighted averages for
8 scattering angles for $dp$- elastic scattering in the horizontal plane only. The typical
values of the beam polarization were $\sim$65-75\% from the ideal values.

\section{The results of the polarization measurements}

\begin{figure}
		\begin{center}
			\includegraphics[width=18pc]{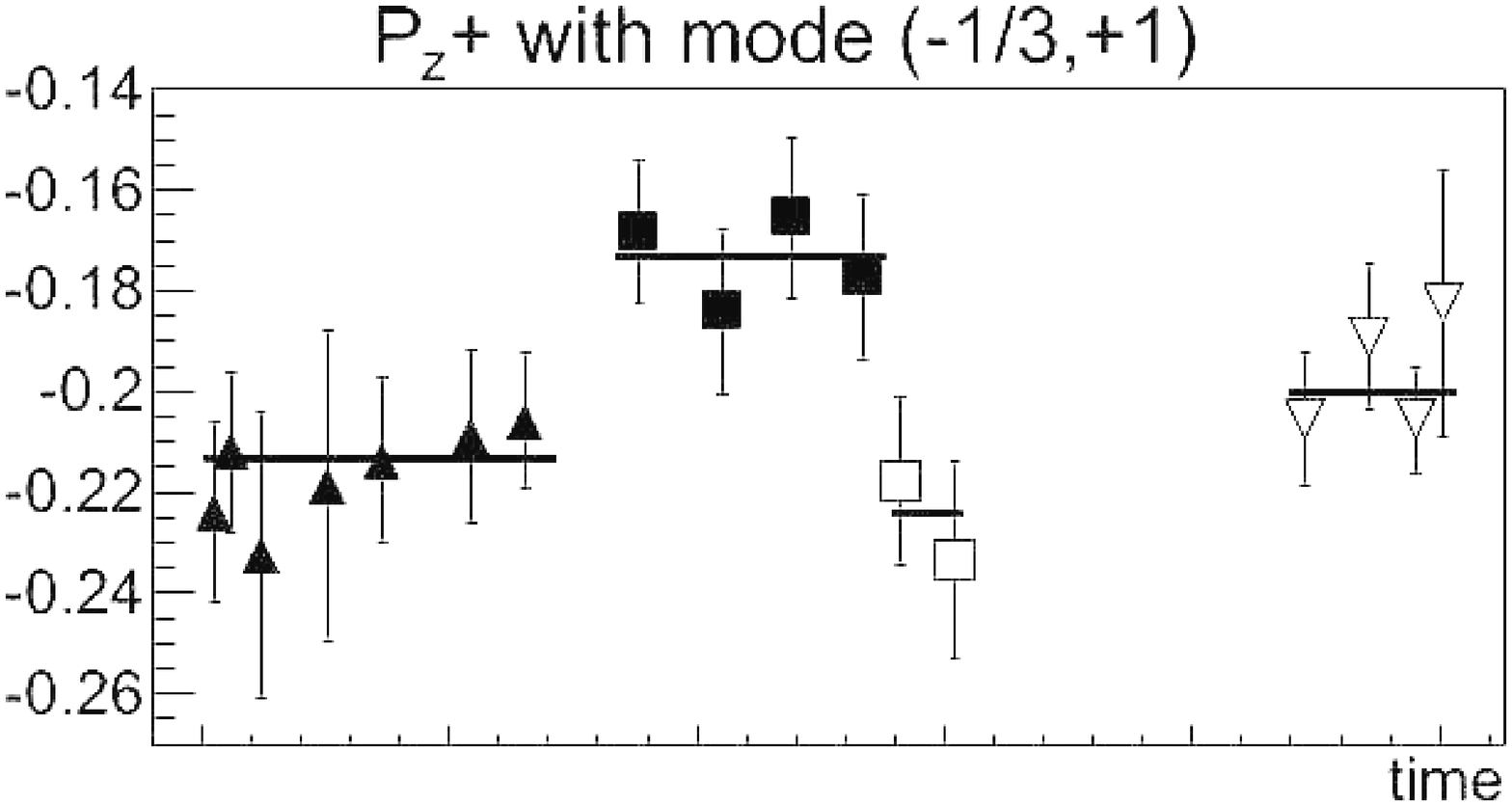}
			\includegraphics[width=18pc]{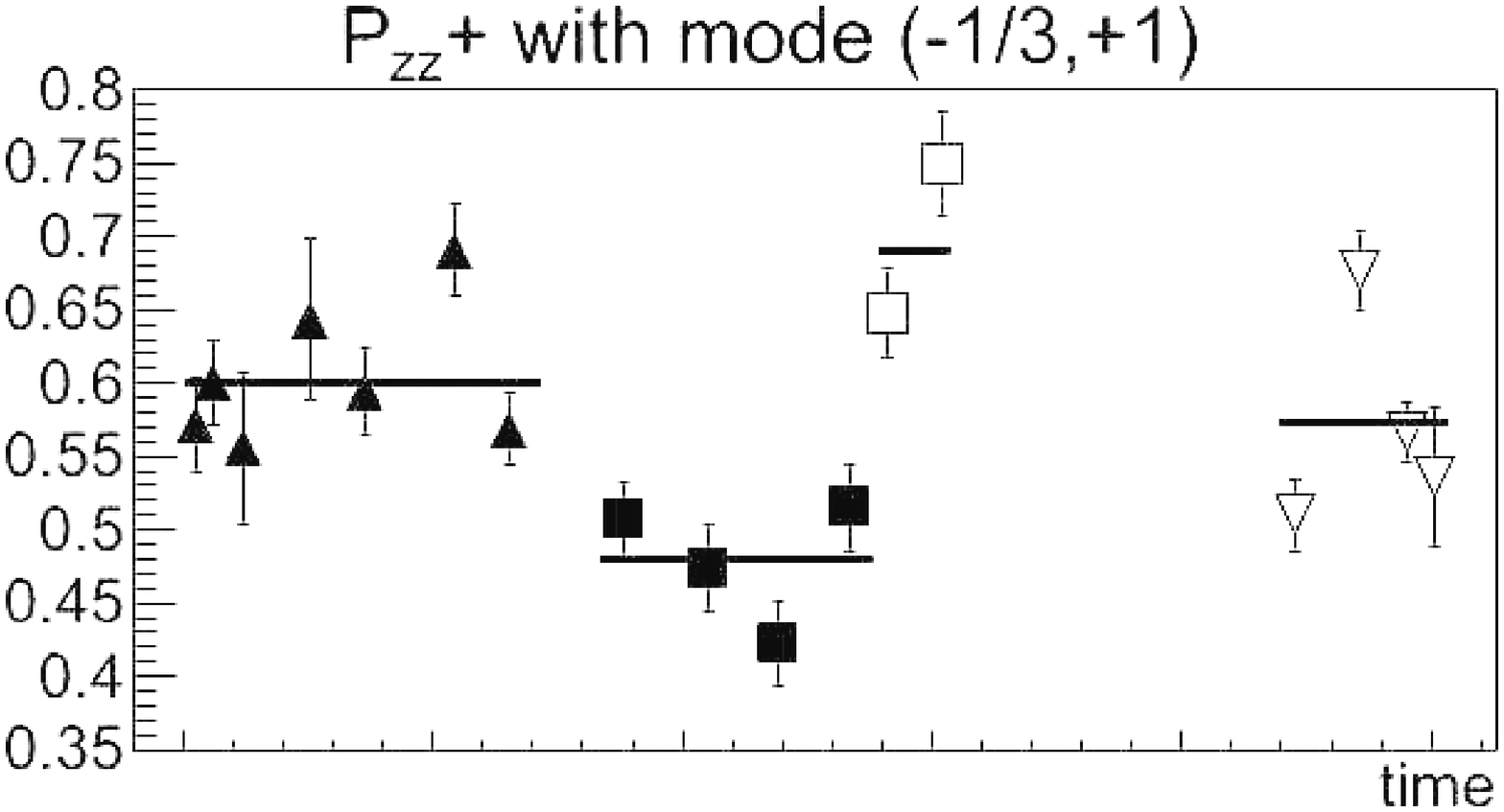}
			\includegraphics[width=18pc]{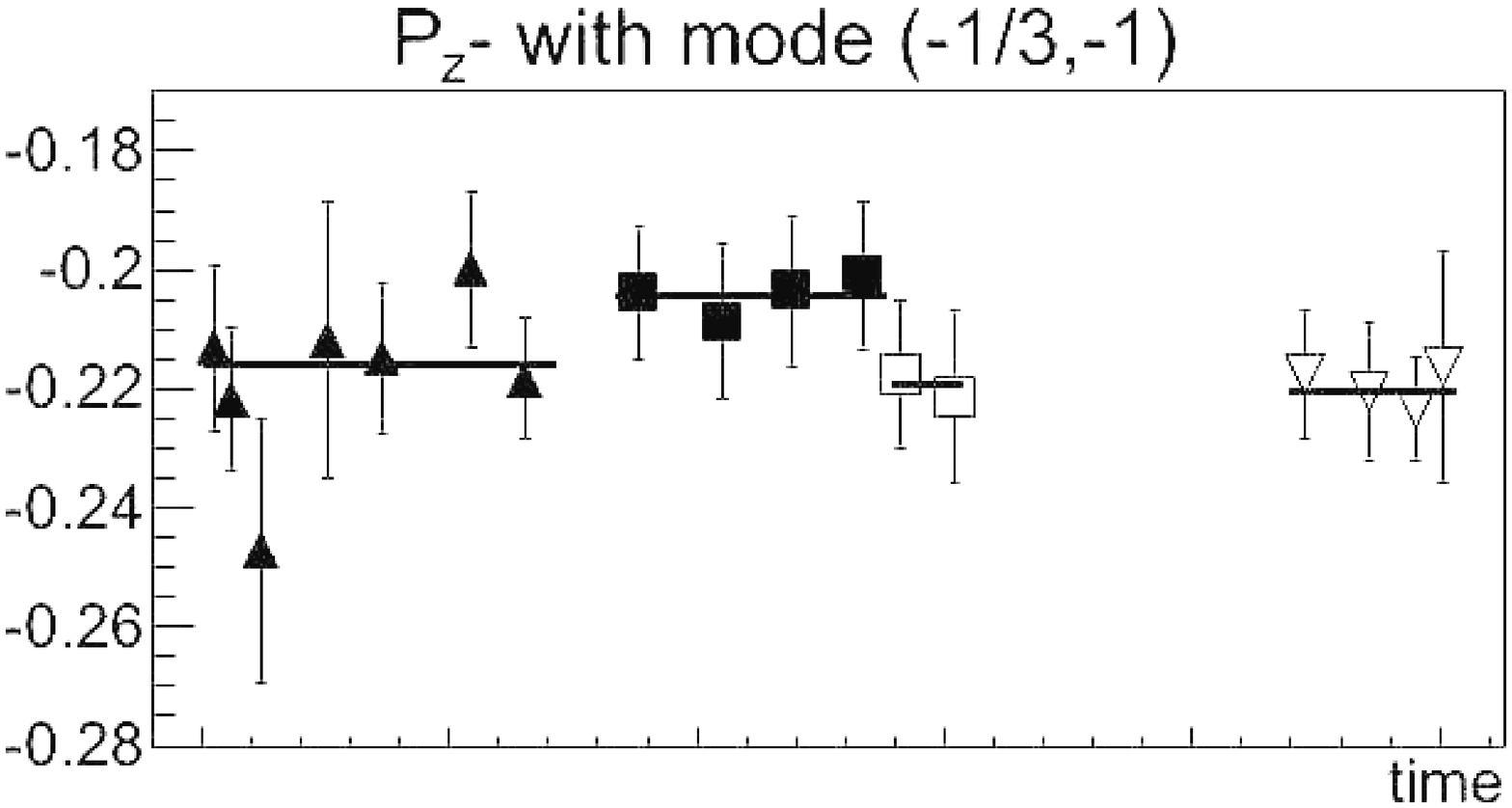}
			\includegraphics[width=18pc]{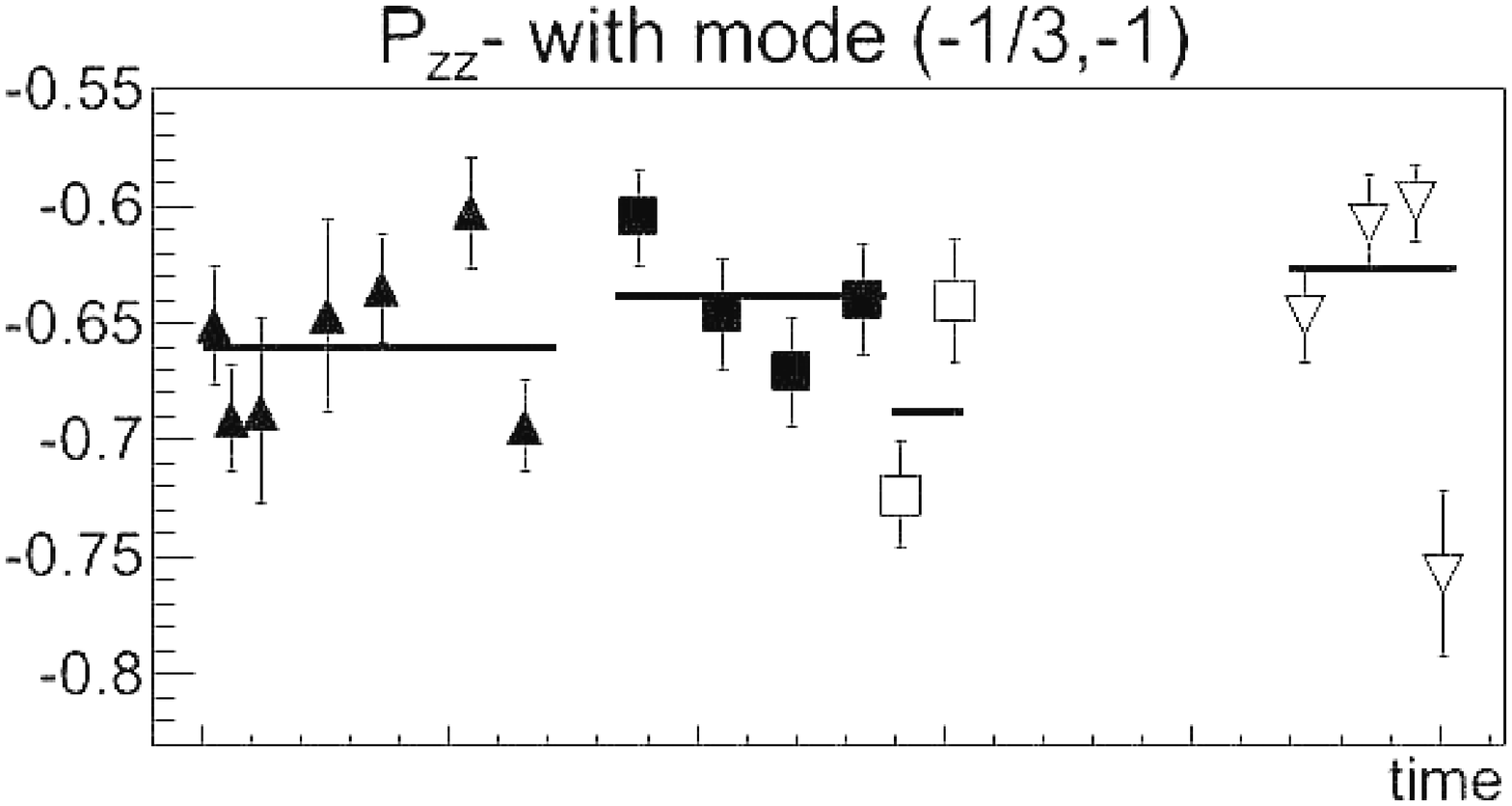}
		\end{center}
		\caption{\label{label}Polarizations values $P_z$ and $P_{zz}$ for spin modes ($P_z$,$P_{zz}$)= (-1/3,-1) and (-1/3, +1)
during the runs in 2016 ~--- 2017 yy.}
        \label{fig:fig1}
	\end{figure}

The vector and tensor polarizations were measured seven, six  and four times  in the parts at November-2016, at December-2016 and at February-2017,
respectively.  The values have  small statistical and systematics errors. They are rather stable within each part of the experiment. 
The  exception is the December-2016 part, when the physics program was separated on two
parts by the tuning of the SPI for pure tensor modes ($P_z$,$P_{zz}$)= (0,-2) and (0,+1)
during 8 hours.   
This is the reason why December-2016 part is divided by sets of the measurements.
The polarization values were approximated by the constants for all four sets of the data.
The results of the measurements and approximation are presented in Fig. \ref{fig:fig1}. 
All the results are  within two standard deviations from these constants.
One can see that the beam polarization values are quite stable within more than  200 hour of the SPI operation. On the other hand, 
SPI demonstrates good reproducibility of the polarization values 
for different sets of the  data after long interruptions.
The typical values of the vector and tensor components of the beam polarization 
for the spin modes (-1/3,-1) and (-1/3,+1) are given in the table~\ref{tab:tab1}.

	\begin{table}
		\caption{\label{label}The vector and tensor polarization for different spin modes of SPI \cite{NewPIS}.}
		\begin{center}
			\begin{tabular}{lllll}
				\br
				Spin mode($P_z$, $P_{zz}$)&$P_z$&$dP_z$&$P_{zz}$&$dP_{zz}$\\
				\mr
                                (-1/3,+1)&-0.254&0.022&0.637&0.039\\
				(-1/3,-1)&-0.223&0.017&-0.621&0.030\\
				(-2/3,+1)&-0.489&0.026&0.631&0.045\\
				(+2/3,0)&0.427&0.021&0.061&0.037\\
				(0,+1)&0.030&0.027&0.880&0.049\\
				(0,-2)&0.046&0.015&-1.469&0.031\\
				\br
			\end{tabular}
        \label{tab:tab1}
		\end{center}
	\end{table}

The polarimeter \cite{ITS_polarimeter} has been used also to tune the SPI \cite{NewPIS} 
operation for pure tensor spin modes (0,-2) and (0,+1), for pure vector spin mode (+2/3,0) and for the  spin mode (-2/3,+1) with both vector and tensor components. The preliminary results are presented in 
table ~\ref{tab:tab1}.
One can see, that the typical
values of the beam polarization were $\sim$65-75\% from the ideal values
for all 6 spin modes of SPI. 

\section{Conclusion}
	The upgraded version of the 270 MeV deuteron beam polarimeter has been used to obtain the vector and tensor polarization during 2016/2017 runs.

	The long-term stability of the vector and tensor components of the beam polarization has been demonstrated for the spin modes (-1/3,+1) and (-1/3,-1) of SPI.

	The polarimeter has been used for tuning of the polarized ion source parameters for 6 different spin modes. 

\ack
The authors thank the Nuclotron staff for providing good conditions of the experiment. They 
thank  V.B.~Shutov  for the help with the SPI \cite{NewPIS} tuning. They express the gratitude to
S.N.~Bazylev, V.I.~Maximenkova, I.V.~Slepnev, V.M.~Slepnev and  A.V.~Shutov for the help 
during the preparation of the detector and DAQ system.
The work has been supported in part by the RFBR under grant $N^0$16-02-00203a,
JINR- Slovak Republic and JINR-Romania scientific cooperation programs in 2016-2017.

\end{document}